\documentclass[eqsecnum,showpacs,amsmath,amssymb,superscriptaddress,nofootinbib,preprintnumbers]{revtex4-2}

\usepackage{graphicx}
\usepackage{dcolumn}
\usepackage{bm}
\usepackage{color}

\newcommand{\bea}{\begin{eqnarray}}
\newcommand{\eea}{\end{eqnarray}}
\newcommand{\vs}[1]{\vspace{#1 mm}}

\newcommand{\hide}[1]{}

\begin{document}

\preprint{NITEP 276}

\title{
Bianchi-I Cosmology with Radiation in Asymptotically Safe Gravity 
}

\author{Chiang-Mei Chen}
\email{cmchen@phy.ncu.edu.tw}
\affiliation{Department of Physics, National Central University, Zhongli, Taoyuan 320317, Taiwan}
\affiliation{Center for High Energy and High Field Physics (CHiP),
National Central University, Zhongli, Taoyuan 320317, Taiwan}
\affiliation{Department of Physics and Kobayashi-Maskawa Institute, Nagoya University, Nagoya 464-8602, Japan}
\affiliation{Department of Physics, Rikkyo University, Toshima, Tokyo 171-8501, Japan}

\author{Ting-Kui Fan}
\email{109202525@cc.ncu.edu.tw}
\affiliation{Department of Physics, National Central University, Zhongli, Taoyuan 320317, Taiwan}

\author{Rituparna Mandal} \email{drimit.ritu@gmail.com}
\affiliation{Department of Physics, National Central University, Zhongli, Taoyuan 320317, Taiwan}

\author{Nobuyoshi Ohta} \email{ohtan.gm@gmail.com}
\affiliation{Nambu Yoichiro Institute of Theoretical and Experimental Physics (NITEP),
Osaka Metropolitan University, Osaka 558-8585, Japan}
\affiliation{Department of Physics and Astronomy, Kwansei Gakuin University, Sanda, Hyogo 669-1330, Japan}

\date{\today}

\begin{abstract}
We study the late-time evolution of an anisotropic Bianchi-I universe with radiation in the framework of asymptotically safe gravity.
We first discuss the radiation-dominated universe for the perfect fluid with the equation of state $p=\rho/3$, and find that the classical evolution involves logarithmic terms, which lead to a slow approach toward isotropy. The quantum effects introduce subleading corrections that soften the anisotropy in the intermediate stage.
Next we discuss the universe with magnetic fields.
For a vanishing classical cosmological constant, we find that the universe in general evolves toward a Kasner-type regime with persistent anisotropy while the expansion rate is enhanced by quantum effects, leading to a faster decay of the magnetic field. In contrast, for a nonzero classical cosmological constant, the late-time dynamics are dominated by the cosmological constant, and the universe asymptotically approaches an isotropic de Sitter phase with exponential decay of both anisotropies and the magnetic field.
Finally, we employ Hodge duality to demonstrate that these cosmological findings apply equally to environments dominated by electric fields.

\end{abstract}

\maketitle

\section{Introduction}
\label{introduction}
Quantum gravity effects play an important role in exploring the universe and addressing several fundamental questions in cosmology. In recent years, the functional renormalization group (RG) approach called the asymptotic safety program~\cite{Weinberg, Wetterich:1992yh, Morris:1993qb, Reuter:1996cp, Souma:1999at, Reuter:2003ca, Percacci:2017, Eichhorn:2018yfc, Reuter:2019book} has emerged as a promising candidate for constructing a consistent and predictive quantum theory of gravity. In particular, it offers a unified description of the universe’s evolution, ranging from the very early epoch to late-time cosmological dynamics~\cite{Bonanno:2001xi, Bonnano:2002plb, BONNANO20041, Bentivegna_2004, Reuter:2005jcap, Bonanno:2007wg, BF_2017, lit1, Platania_2020, Alvarez2022, Alvarez2024, Zholdasbek:2025, Chen:2024ebb, Chen:2025ybu}. The key feature underlying this framework is the existence of a non-trivial ultraviolet (UV) fixed point of the RG flow. This property ensures that the theory remains non-perturbatively renormalizable~\cite{Baldazzi:2021orb, Ohta:2025xxo}, in contrast to perturbative approaches to quantum gravity. Moreover, the presence of such a fixed point implies that the RG trajectories of all relevant couplings are confined to a finite-dimensional critical surface and approach the fixed point in the UV limit, thereby guaranteeing predictability.

To study asymptotic safety, the functional RG technique serves as a powerful tool for analyzing the RG flow of the effective average action, $\Gamma_{k}[g_{\mu\nu}]$. This action incorporates quantum fluctuations above a nonzero momentum cutoff scale $k$. As a function of the RG scale, $\Gamma_{k}[g_{\mu\nu}]$ evolves along trajectories determined by an exact RG equation. This equation defines a Wilsonian flow on the space of diffeomorphism-invariant functionals of the metric $g_{\mu\nu}$ and interpolates between the bare action in the UV limit $k \to \infty$ and the full quantum effective action in the infrared (IR) limit $k \to 0$. The flow equation governing this evolution is given by
\begin{equation}
k \partial_k \Gamma_k = \frac12 {\rm Tr} \left\{ \left[ \Gamma_k^{(2)} + R_k(\Delta) \right]^{-1} k \partial_k R_k(\Delta) \right\}\,,
\label{frge}
\end{equation}
where $\Gamma_k^{(2)}$ denotes the second functional derivative of $\Gamma_k$ with respect to the fluctuation field, evaluated at a fixed background. The regulator function $R_k(\Delta)$ implements an IR cutoff via an appropriate differential operator $\Delta$, as extensively discussed in~\cite{Litim:2001, Codello:2008vh}.

The construction of the flow equation is based on the background field formalism, in which the full metric is decomposed into a background part and fluctuations around it. The regulator term suppresses contributions from modes with eigenvalues $p^{2} \lesssim k^2$ by effectively introducing a scale-dependent mass term. Modes with $p^{2} \gtrsim k^2$ remain unsuppressed, as the regulator vanishes in this regime. Consequently, high-momentum modes are integrated out in the usual manner, ensuring a smooth interpolation between microscopic (UV) and macroscopic (IR) physics.

Approximate non-perturbative solutions of the RG equation can be obtained by truncating the action to a finite-dimensional subspace that captures the relevant physical degrees of freedom. The simplest realization of this approach in gravity is the Einstein–Hilbert truncation, where the RG flow is confined to a two-dimensional subspace spanned by the operators $\int d^4 x \sqrt{-g}\,R(g_{\mu \nu})$ and $\int d^4 x \sqrt{-g}$. Although the present analysis is performed within the Einstein–Hilbert truncation, one can always extend the truncation ansatz to include higher-derivative curvature invariants. Nevertheless, studies of higher-derivative truncations have shown that the qualitative structure of the renormalization group flow near the non-Gaussian fixed point is already well captured within the Einstein–Hilbert truncation \cite{Lauscher:2002,Codello:2006in,deBerredo-Peixoto:2004cjk,Ohta:2013uca}. In particular, the existence and stability properties of the fixed point remain robust under the inclusion of additional higher-curvature operators, although quantitative details of the flow may be modified.
	
In the Einstein–Hilbert truncation, the associated scale-dependent couplings are the Newton gravitational coupling $G(k)$ and the cosmological term $\Lambda(k)$, which appear as prefactors of these terms. Their scale dependence is captured by the beta functions, which are the system of ordinary differential equations obtained by inserting the truncation ansatz into the RG flow equation. For applications to late-time cosmology, corresponding to the infrared regime of the RG flow, the running couplings can be expanded as a perturbative series in powers of the dimensionless quantity $G_0k^2$, assuming $G_0k^2 < 1$. In this regime, the running couplings are well approximated by the first few terms in the infrared expansion. The present framework therefore, describes effective late-time quantum-gravitational corrections rather than the ultraviolet Planck-scale regime. The corresponding infrared expansions are given by~\cite{Chen:2024ebb}
\begin{eqnarray}
\label{eq_GLam}
G(k) &=& G_0 \left[ 1 - \omega G_0 k^2 + \omega_1 G_0^2 k^4 + \mathcal{O}\left( G_0^3 k^6 \right) \right],
\nonumber\\
\Lambda(k) &=& \Lambda_0 \left[ 1 - \mu G_0 k^2 + \mu_1 G_0^2 k^4 + \mathcal{O}\left( G_0^3 k^6 \right) \right]
+ G_0 k^4 \left[ \nu + \nu_1 G_0 k^2 + \mathcal{O}(G_0^2 k^4) \right],
\label{gravterms}
\end{eqnarray}
where $G_0$ and $\Lambda_0$ are the classical Newton and cosmological constants, respectively,
and the dimensionless parameters $\omega, \omega_1, \mu, \mu_1, \nu$ and $\nu_1$ are given, for $\Lambda_0 = 0$, by
\begin{equation}
\omega = \frac{11}{6 \pi}, \quad
\omega_1 = \frac{217}{72 \pi^2}, \quad
\nu = \frac1{8 \pi}, \quad
\nu_1 = \frac{7}{54 \pi^2},
\label{data1}
\end{equation}
and for $\Lambda_0 \ne 0$, they are
\begin{eqnarray}
\omega = \mu = \frac{7}{6 \pi},
\quad \omega_1 = \frac{49}{36 \pi^2}  - \frac{5}{24 \pi G_0 \Lambda_0},
\quad \mu_1 = \frac{49}{36 \pi^2}, \quad
\nu = - \frac{17}{24 \pi}, \quad
\nu_1=\frac{119}{144\pi^2}.
\label{data2}
\end{eqnarray}
Note that the cosmological term $\Lambda$ vanishes only for $\Lambda_0=0$ in the classical case $k=0$; the quantum corrections induce nonzero cosmological term even when the classical cosmological constant $\Lambda_0$ is zero.

The radiation-dominated universe is important in the history of our universe. The most studied case is the perfect fluid of homogeneous and isotropic radiation with the equation of state $p = \rho/3$. More generally, we expect that the presence of magnetic fields may introduce a preferred spatial direction (see reviews~\cite{Giovannini2004, Durrer2013} and references cited therein). Such fields can leave observable imprints on the anisotropy of the cosmic microwave background, depending on the orientation of the field lines~\cite{Jacobs:1969, Madsen1989}. This requires a spacetime description more general than the isotropic Friedmann-Lemaître-Robertson-Walker (FLRW) model. Anisotropic cosmological models, such as those of Bianchi type, provide a natural framework for investigating the dynamical effects of magnetic fields. In particular, the Bianchi-I (BI) universe is of special importance, as it represents the simplest anisotropic generalization of the FLRW spacetime while retaining spatial homogeneity, and enables the study of deviations from isotropy as well as the origin of large-scale structures beyond standard Friedmann models~\cite{Jacobs:1969, Fabbri1987}. Anisotropic cosmological models also provide a natural framework for investigating possible quantum-gravitational effects in cosmology. The scale-dependent gravitational couplings can modify the anisotropic expansion and influence the evolution toward isotropy. In this context, BI cosmologies offer a useful setting for studying how quantum corrections affect anisotropic dynamics and the evolution of large-scale magnetic fields.
The homogeneous but anisotropic BI cosmological model we consider in this paper has the line element
\begin{equation}
\label{Bianchi_I}
ds^2 = - dt^2 + a_1^2(t) dx^2 + a_2^2(t) dy^2 + a_3^2(t) dz^2,
\end{equation}
where $a_1(t), a_2(t)$ and $a_3(t)$ denote the directional scale factors, each depending on the cosmic time $t$. For this metric, the nonvanishing components of the Ricci tensor and the Ricci scalar are obtained as
\begin{equation}
R_{00} = -3 \dot{H} - \sum_{i=1}^3 H_i^2,
\qquad
R_{ii} = a_i^2 \left( \dot{H}_i + 3 H H_i \right),
\qquad
R = 6 \dot{H} + 9 H^2 + \sum_{i=1}^3 H_i^2,
\end{equation}
where $H_i$ are the directional Hubble parameters, and $H$ represents their mean value, defined as
\begin{equation}
H_i = \frac{\dot{a}_i}{a_i},
\qquad
H = \frac{1}{3} \sum_{i=1}^3 H_i.
\end{equation}

Classical studies, including the exact vacuum solution by Kasner~\cite{Kasher:1921zz} and the Heckmann–Schücking solution for dust~\cite{Heckmann:1959}, along with its generalizations~\cite{Khalatnikov:2003ph, Kamenshchik:2009dt}, have been instrumental in understanding cosmological singularities~\cite{Lifshitz:1963, Belinskii:1970} and the isotropization of the BI universe in the presence of matter~\cite{Misner:1967uu, Jacobs:1968}. A considerable body of work has explored BI cosmologies from various perspectives~\cite{Hu:1978zd, Wald:1983ky, Cho:1995hz, Chen:2000gaa, Saha:2001ig, Saha:2006iu, Russell:2013oda, Jacobs:1969, Rosen:1962, Rosen:1964, Doroshkevich:1965, Thorne:1967, Tupper:1977c, Singh1989}. These analyses have been further extended to other Bianchi classes, including types II, III, VI, VIII and IV~\cite{Hughston:1970, Dunn:1976, Tupper:1977, Lorenz:1980pla, Dunn:1980, Lorenz:1980b}. 

On the other hand, substantial body of literature has also explored anisotropic, spatially homogeneous universes with large-scale magnetic fields in various contexts~\cite{Salim1998, Tsagas2000, Horwood:2004, Bronnikov:2004nu, Barrow:2007, King2007, Watanabe2009, Yadav:2012, Liu2017, Gioia2019, Tarai2020, Casadio:2023ggm, Muharlyamov:2025}.
Despite these developments, studies incorporating quantum gravitational effects remain relatively scarce. In particular, only a limited number of works have examined anisotropic settings with homogeneous magnetic fields within the framework of effective loop quantum cosmology (LQC)~\cite{Maartens:2008, Rikhvitsky:2014, Motaharfar:2024}. In effective LQC, quantum corrections are incorporated by modifying the classical Hamiltonian with terms motivated by the underlying discrete quantum geometry of LQG. The resulting quantum-improved effective Hamiltonian then leads to modified cosmological equations, which can replace the classical singularity by a nonsingular quantum bounce \cite{Bojowald:2005, Ashtekar:2009, Ashtekar:2011}. In magnetized Bianchi-I models, quantum corrections can significantly modify the evolution of anisotropy and magnetic fields, leading to nonsingular bouncing cosmologies and nontrivial shear dynamics~\cite{Maartens:2008, Rikhvitsky:2014, Motaharfar:2024}. Similarly, BI anisotropic magnetized cosmological models with a bulk viscous fluid and a time-dependent cosmological ``constant'' $\Lambda$ have been investigated, typically under the assumption that $\Lambda$ decreases with cosmic time~\cite{Pradhan:2003, Pradhan2004}. In contrast, the present asymptotic safety framework incorporates quantum effects through the renormalization group running of the gravitational couplings $G(k)$ and $\Lambda(k)$, which are determined by the underlying renormalization group flow rather than being introduced phenomenologically or through effective quantum geometry corrections as in LQC.

In our previous paper~\cite{Chen:2025ybu}, we investigated the dynamics of a BI universe filled with a perfect fluid, first at the classical level and subsequently including quantum corrections to the gravitational coupling and the cosmological term within the asymptotic safety framework. We observed that, in the case of a radiation-dominated universe with the equation of state $p = \rho/3$, a special class of solutions arises when $\Lambda_0 = 0$, characterized by the presence of logarithmic terms. These terms do not appear for other values of the equation-of-state parameter $w$, and they significantly complicate the analysis, making the general power-series expansion singular in the radiation-dominated limit. For this reason, the radiation-dominated case was not examined in detail in our previous general study of BI cosmologies~\cite{Chen:2025ybu}. Nevertheless, given the central role of the radiation-dominated era in cosmic evolution, this special case may have important implications for cosmological history.  In particular the quantum effects in this case are left largely unexplored in the literature. This motivates us to investigate in detail how asymptotically safe quantum-gravitational corrections modify the logarithmic late-time behavior, anisotropic evolution, and isotropization properties of these radiation-dominated BI solutions.

In the present work, we perform a detailed study of radiation-dominated BI cosmology with $\Lambda_0=0$, both at the classical and quantum-improved levels. In particular, we derive the corresponding power-series solutions in cosmic time and explicitly analyze how the logarithmic contributions appearing in the solutions influence the evolution toward isotropy, both in the classical case and in the presence of quantum-gravitational corrections.	
We then incorporate quantum corrections directly at the level of the Einstein equations by promoting the classical Newton coupling and cosmological constant to scale-dependent quantities, as specified in~\eqref{eq_GLam} -- \eqref{data2}, rather than modifying the gravitational action or the matter field equations~\cite{Reuter:2003ca}. In approaches based on modifications of the gravitational action, such as Brans–Dicke-type formulations, the scale-dependent couplings $G$ and $\Lambda$  effectively behave as spacetime-dependent fields and contribute to the energy-momentum content of the system, leading to nontrivial consistency conditions and restrictions on the possible cutoff identifications. In the present work, we instead adopt an RG-improved equation-level approach, which effectively captures the leading-order quantum-gravitational corrections while avoiding these additional complications. A crucial step in this procedure is the identification of the renormalization group (RG) scale $k$ with a physically meaningful, time-dependent quantity. Following our earlier work~\cite{Chen:2024ebb, Chen:2025ybu}, we identify the RG scale $k$ with the quantum Hubble parameter.

Another purpose of this paper is to examine a configuration in which a strong magnetic field is aligned along a single spatial direction. We find that, once quantum corrections are incorporated, the Einstein equations become overdetermined, leading to an inconsistency in the system. To restore consistency, we introduce an additional quantum-induced energy density. We then construct power-series solutions in cosmic time for both $\Lambda_0 = 0$ and $\Lambda_0 > 0$. For $\Lambda_0 = 0$, the late-time behavior of the subleading terms is governed by the anisotropy parameter, resulting in a Kasner-type universe. In contrast, for $\Lambda_0 > 0$, the late-time dynamics are dominated by the classical cosmological constant, while the magnetic field strength decays exponentially.

We further analyze the impact of quantum corrections on the isotropization process in the presence of a magnetic field. Our results indicate that, even with the inclusion of an additional quantum-induced energy density, the BI universe does not dynamically approach an FLRW geometry at intermediate stages. Nevertheless, for $\Lambda_0 > 0$, isotropy is asymptotically achieved at late times, in agreement with the cosmic no-hair theorem, and the universe evolves toward a de Sitter state. This analysis also provides insight into the initial magnetic field strength inferred from the late-time behavior when quantum corrections are taken into account.

The paper is organized as follows. In Sec.~II, we present the quantum-improved Einstein equations for a BI universe filled with a perfect fluid satisfying $p_1 = p_2 = p_3 = \rho/3$, corresponding to a radiation-dominated universe. We first discuss the classical evolution for $\Lambda_0 = 0$, and then examine the effects of quantum corrections. We also analyze the conditions under which isotropy may emerge in this setting. This case was not studied in Ref.~\cite{Chen:2025ybu} for simplicity. The case with $\Lambda_0 > 0$ has already been discussed there, and we refer the reader to that work for details. In Sec.~\ref{BImag}, we consider a magnetic field aligned along the $z$-direction and formulate the corresponding Einstein-Maxwell system as a concrete example of the universe filled with the radiation. We show that, once quantum corrections are included, the system becomes overdetermined, necessitating the introduction of an additional traceless energy–momentum tensor to restore consistency. We then analyze the classical BI cosmology and derive the leading-order quantum corrections to the volume element and the quantum-induced energy density for both $\Lambda_0 = 0$ and $\Lambda_0 > 0$ within the asymptotic safety framework. Finally, we investigate how quantum corrections influence the evolution of the magnetic field strength and use the late-time behavior to infer the initial magnetic field strength. In Sec.~\ref{BIele}, we extend the study to the electric fields. We demonstrate the equivalence between electric and magnetic field configurations aligned along a single spatial direction via Hodge duality, including the effects of quantum corrections implemented at the level of the Einstein equations in a cosmological setting. So we can just borrow the results for magnetic fields. Finally, in Sec.~\ref{summary}, we summarize our main results.

\section{Radiation-dominated universe: Perfect-Fluid Matter}

In this section, we consider the BI universe for the perfect fluid. The energy-momentum tensor is assumed to take the diagonal form
\begin{equation} \label{eq_EMT}
T^\mu{}_\nu = \mathrm{diag}(-\rho, p_1, p_2, p_3),
\end{equation}
and to satisfy the traceless condition,
\begin{equation}
T = T^\mu{}_\mu = -\rho + p_1 + p_2 + p_3 = 0.
\end{equation}

The Einstein field equations are written as
\begin{equation}
R_{\mu\nu} = \Lambda g_{\mu\nu} + 8 \pi G \left( T_{\mu\nu} - \tfrac{1}{2} T g_{\mu\nu} \right)
= \Lambda g_{\mu\nu} + 8 \pi G T_{\mu\nu},
\label{emtpf}
\end{equation}
where the last equality follows from the traceless condition $T = 0$.

For the BI metric~\eqref{Bianchi_I} and the energy-momentum tensor~\eqref{eq_EMT}, the Einstein equations decompose into the following set of evolution equations:
\begin{eqnarray}
3 \dot{H} + \sum_{i=1}^3 H_i^2 &=& \Lambda - 8 \pi G \rho,
\label{eq_Ein_BI_tt} \\
\dot{H}_i + 3 H H_i &=& \Lambda + 8 \pi G p_i,
\label{eq_Ein_BI_ii}
\end{eqnarray}
expressed in terms of the directional Hubble parameters $H_i$ and their average $H$.

For a perfect fluid source with isotropic pressure,
\begin{equation}
p_1 = p_2 = p_3 = \rho / 3,
\label{isop}
\end{equation}
the Einstein equations yield, after some algebraic manipulation, for the details see~\cite{Chen:2025ybu}, the following evolution equations for the total volume element $V(t) = a_1(t) a_2(t) a_3(t)$ and the energy density $\rho(t)$:
\begin{eqnarray}
&& \frac{\ddot{V}}{V} - \frac{1}{3} \frac{\dot{V}^2}{V^2} + \frac{1}{2} \frac{\kappa^2}{V^2} = 2 \Lambda,
\label{eq_Ein_BI_V} \\
&& \rho = \frac{1}{8 \pi G} \left( \frac{1}{3} \frac{\dot{V}^2}{V^2} - \frac{1}{2} \frac{\kappa^2}{V^2} - \Lambda \right).
\label{eq_Ein_BI_rho}
\end{eqnarray}
Here the directional Hubble parameters have the relations \begin{equation}
H_i = H + \kappa_i/V,
\end{equation}
with constraint $\sum_{i=1}^3 \kappa_i = 0$ and the anisotropic parameter $\kappa^2$ in the equations is defined as $\kappa^2 = \sum_{i=1}^3 \kappa_i^2$.
The values of $\kappa_i$ are determined by the initial condition. Anisotropy is determined by the initial condition since the radiation has no source of anisotropy, and eventually it decays out according to the expansion of the universe.

We can regard Eq.~\eqref{eq_Ein_BI_V} as the equation determining the total volume $V$ and then Eq.~\eqref{eq_Ein_BI_rho} as that determining the energy density.
Notice that Eq.~\eqref{eq_Ein_BI_V} remains invariant under the scaling transformation for any $\Lambda$:
\bea
{\rm (i)} && \kappa \rightarrow \alpha \, \kappa, \qquad V \rightarrow \alpha \, V.
\label{symm1}
\eea
In addition, for $\Lambda_0=0$ and in the classical case ($k=0$), the total cosmological constant $\Lambda=0$, and
Eq.~\eqref{eq_Ein_BI_V} is also invariant under
\bea
{\rm (ii)} && \kappa \rightarrow \alpha \, \kappa, \qquad t \rightarrow \alpha^{-1} \, t.
\label{symm2}
\eea
These symmetries will later be used to constrain the form of the perturbative solution.
    
\subsection{Classical solution for $\Lambda_0=0$}

Here we see that $\Lambda=0$ according to Eq.~\eqref{gravterms} at the classical level $k=0$.
The solution to Eq.~\eqref{eq_Ein_BI_V} can be constructed as a power series expansion in $\kappa^2$, yielding
\begin{equation}
V(t) = A_0 (t - t_0)^{3/2} + \kappa^2 \frac{X}{2 A_0} (t - t_0)^{1/2} + \kappa^4 \frac{X^2 + 4 X + 6}{24 A_0^3} (t - t_0)^{-1/2} - \kappa^6 \frac{X^3 - 6 X - 6}{432 A_0^5} (t - t_0)^{-3/2} + \mathcal{O}(\kappa^8),
\label{classicalpower}
\end{equation}
where
\begin{equation} \label{eq_def_X}
X = \ln\left[ \frac{A_0^2}{\kappa^2} (t - t_0) \right].
\end{equation}
Here $A_0$ and $t_0$ are integration constants with dimensions $[A_0] = L^{-3/2}$ and $[t_0] = L$, respectively. The argument of the logarithm is dimensionless, as $V(t)$ is dimensionless and $[\kappa] = L^{-1}$. The above solution is constructed so as to remain invariant under the transformations~\eqref{symm1} with $A_0 \to \alpha A_0$ since then both sides are just multiplied by $\alpha$. It is also invariant under \eqref{symm2} with $A_0 \to \alpha^{3/2} A_0$.

This solution is quite different from those found in Ref.~\cite{Chen:2025ybu} in that it contains logarithmic term.
This is a special feature for the radiation-dominated universe.

For $\Lambda = 0$, Eq.~\eqref{eq_Ein_BI_V} can also be rewritten as
\begin{equation}
\frac{d\dot{V}^2}{dV} = \frac{2}{3V} \left( \dot{V}^2 - \frac{3}{2} \kappa^2 \right).
\label{2nde}
\end{equation}
Integration yields
\begin{equation}
\dot{V}^2 = c \, V^{2/3} + \frac{3}{2} \kappa^2,
\end{equation}
where $c$ is an integration constant. The corresponding implicit solution is given by
\begin{equation}
t - t_0 = \int_{V_0}^{V} \frac{dV}{\sqrt{c \, V^{2/3} + 3 \kappa^2/2}}.
\end{equation}
The integral on the right-hand side can be performed analytically, giving
\begin{equation}
\frac23 c^{3/2} (t - t_0) = \left[ \sqrt{c} V^{1/3} \sqrt{c V^{2/3} + 3 \kappa^2/2} - \frac{3 \kappa^2}2 \ln\left( \sqrt{c} V^{1/3} + \sqrt{c V^{2/3} + 3 \kappa^2/2} \right) \right]_{V_0}^V.
\label{intsol}
\end{equation}
This gives the exact solution if we could solve for $V$. Thus the form~\eqref{2nde} is useful for obtaining exact solution even though it is sometimes difficult to give explicit solutions in terms of elementary functions, whereas Eq.~\eqref{eq_Ein_BI_V} is useful for finding approximate power solutions.
Equation~\eqref{intsol} contains a logarithmic term, distinguishing the radiation-dominated case ($w = 1/3$) from other values of $w$~\cite{Chen:2025ybu}.
It is straightforward to verify that, in the isotropic limit $\kappa = 0$, the solution reduces to~\eqref{classicalpower} evaluated at $\kappa = 0$.

The appearance of the logarithmic term in the solution~\eqref{classicalpower} reflects the unique behavior of the radiation-dominated universe ($w = 1/3$) within the BI framework. In this case, the anisotropy parameter $\kappa$ couples to the expansion dynamics in a nontrivial way, leading to a slower isotropization rate compared to the matter- or vacuum-dominated eras. The logarithmic correction thus encodes the residual influence of anisotropy on the expansion of the universe, even as $V(t)$ grows with time. This feature is absent for other values of $w$, underscoring the special role of the radiation-dominated epoch in connecting anisotropic early-universe dynamics to the subsequent isotropic evolution.

\subsection{Quantum-improved solution with perfect-fluid matter and $\Lambda_0 = 0$}

Here we incorporate the quantum effects at the ``equation level” by replacing the ordinary Newton coupling and cosmological constant in the Einstein equations with scale-dependent (running) coupling parameters~\eqref{gravterms}.
We solve the Einstein equations~\eqref{eq_Ein_BI_V} and~\eqref{eq_Ein_BI_rho} with these couplings $G(k)$ and $\Lambda(k)$. A crucial step is to identify the RG scale $k$ with a physically meaningful, time-dependent quantity.

Following Ref.~\cite{Chen:2025ybu}, and motivated by the typical identification of the RG scale with the Hubble parameter, we adopt the relation
\begin{equation}
k^2 = \xi^2 \frac{\dot V_{q}^2}{V_{q}^2} = \xi^2 \left( c V_{q}^{-4/3} + \frac32 \kappa^2 V_{q}^{-2} \right),
\end{equation}
where $V_q$ is a quantum-improved volume and $\xi$ is a dimensionless constant of order unity. This scale identification induces quantum corrections to the cosmological term through the low-energy expansion of the functional RG flow equation~\cite{Chen:2024ebb}, given in Eq.~\eqref{gravterms}.
For the optimized cutoff scheme~\cite{Litim:2001}, the quantum coefficients take the values $\nu = 1/8\pi, \, \nu_1 = 7/54 \pi^2$. Substituting the above scale relation yields the explicit volume dependence
\begin{equation}
\Lambda(V_{q}) = G_0 \tilde\nu \left( c V_{q}^{-4/3} + \frac{3}{2} \kappa^2 V_{q}^{-2} \right)^2 + G_0^2 \tilde\nu_1 \left( c V_{q}^{-4/3} + \frac{3}{2} \kappa^2 V_{q}^{-2} \right)^3 + \cdots,
\end{equation}
where $\tilde\nu = \xi^4 \nu, \tilde\nu_1 = \xi^6 \nu_1$. 

Retaining only the leading quantum correction proportional to $\tilde\nu$, the quantum-improved solution for the total volume is found to be
\begin{eqnarray}
V_{q}(t) &=& \left[ A_0 (t - t_0)^{3/2} + \frac{G_0 c^2 \tilde\nu}{A_0^{5/3}} (t - t_0)^{-1/2} + \cdots \right]
\nonumber\\
&+& \kappa^2 \left[ \frac{X}{2 A_0} (t - t_0)^{1/2} + \left( \frac{G_0 c \tilde\nu}{A_0^{7/3}} - \frac{G_0 c^2 \tilde\nu (X + 1)}{6 A_0^{11/3}} \right) (t - t_0)^{-3/2} + \cdots \right]
\nonumber\\
&+& \kappa^4 \left[ \frac{X^2 + 4 X + 6}{24 A_0^3} (t - t_0)^{-1/2} + \left( \frac{3 G_0 \tilde\nu}{8 A_0^3} - \frac{G_0 c \tilde\nu (3 X + 2)}{6 A_0^{13/3}} + \frac{G_0 c^2 \tilde\nu (3 X^2 + 2 X - 2)}{72 A_0^{17/3}} \right) (t - t_0)^{-5/2} + \cdots \right]
\nonumber\\
&+& \cdots .
\end{eqnarray}
where $X$ is defined in~\eqref{eq_def_X}.
\vs{3}

\noindent
{\bf Physical meaning of the quantum effects:}

The leading quantum correction appears at order $(t - t_0)^{-1/2}$, which corresponds to the $\kappa^4$ term in the classical expansion~\eqref{classicalpower}. This result indicates that the quantum effects are effectively higher-order corrections in  anisotropy, modifying the behavior of the BI universe in the intermediate stage.
The anisotropic universe eventually evolves into an isotropic FLRW universe, as at very late times the leading classical terms govern the dynamics, while the quantum corrections lead to faster isotropization compared to the classical case.

\subsection{Classical and Quantum-improved solutions with perfect-fluid matter for $\Lambda_0 > 0$}

The BI cosmology, both classical and quantum, with $\Lambda_0>0$ has been discussed in~\cite{Chen:2025ybu} with a perfect fluid source. Unlike the case with $\Lambda_0=0$, a radiation source with equation of state $w = 1/3$ leads to logarithmic contributions when we consider a power-series expansion of the solution and therefore it requires a separate analysis. The case with $\Lambda_0>0$ admits a general solution for all values of $w$ (physically in the range $-1 < w \le 1$). A detailed analysis has been presented in~\cite{Chen:2025ybu}. We refer to this paper for the solutions and further discussions on this case.

\section{BI universe with magnetic field}
\label{BImag}

Instead of the perfect fluid model, we consider the Einstein–Maxwell theory with a cosmological term, described by the action
\begin{equation}
S = \int d^4x \sqrt{-g} \left( \frac{R - 2 \Lambda}{16 \pi G} - \frac{1}{4 \alpha} F_{\mu\nu} F^{\mu\nu} \right),
\qquad
F_{\mu\nu} = \partial_\mu A_\nu - \partial_\nu A_\mu,
\end{equation}
where $\alpha$ denotes the Maxwell coupling.
Varying the action with respect to the metric and the gauge field leads to the Einstein and Maxwell equations:
\begin{eqnarray}
&& R_{\mu\nu} - \frac12 R g_{\mu\nu}
= - \Lambda g_{\mu\nu} + \frac{8 \pi G}{\alpha} \left( F_{\mu\alpha} F_\nu{}^\alpha - \frac14 g_{\mu\nu} F_{\alpha\beta} F^{\alpha\beta} \right)
\equiv - \Lambda g_{\mu\nu} + 8 \pi G \, T_M{}_{\mu\nu}, 
\label{eq_Einstein} \\
&& \nabla_\mu F^{\mu\nu} = 0, \qquad \partial_{[\mu} F_{\nu\lambda]} = 0.
\label{eq_Maxwell}
\end{eqnarray}
where $T_M{}_{\mu\nu}$ represents the energy-momentum tensor of the electromagnetic field. Here we include quantum corrections to the equations through the coupling constants. However, since we have not considered any interaction between photons and charged particles, the Maxwell coupling $\alpha$ does not receive quantum corrections from the matter. It is true that gravity may induce quantum corrections to $\alpha$ but these effects are expected to be very small; therefore, we absorb it as a constant in the energy–momentum tensor.

Let us consider the case with a purely spatially homogeneous (constant) magnetic field aligned along the $z$--direction\footnote{Only a single spatial component of the gauge field can be consistently activated; otherwise, off-diagonal components of the energy-momentum tensor would be generated, which are incompatible with the BI geometry. A perfect fluid corresponds to the situation in which the electromagnetic field has no preferred spatial direction. In that case, one should take the expectation value (or average) of the energy–momentum tensors for electromagnetic fields oriented along the $x$-, $y$- and $z$-directions, giving the form of the perfect fluid considered in the preceding section.}
\begin{equation}
F_{21} = - F_{12} = B_z.
\label{magnetic}
\end{equation}
For the BI spacetime, the energy-momentum tensor can be computed from $T_M{}_{\mu\nu}$, defined in Eq.~\eqref{eq_Einstein} in terms of the electromagnetic field tensor. The energy-momentum tensor of the electromagnetic field then has the nonvanishing components
\begin{equation}
\label{eq_T_M}
T_M{}^\mu{}_\nu = {\rm diag}(-\rho_M, \rho_M, \rho_M, -\rho_M), \qquad \rho_M = \frac{B_z^2}{2 \alpha a_1^2 a_2^2}~.
\end{equation}
Compared with the traceless isotropic energy-momentum tensor for the perfect fluid with~\eqref{emtpf} and \eqref{isop}, this is still traceless though anisotropic.
The Einstein equations~\eqref{eq_Einstein} reduce to the following set of equations, corresponding respectively to the $(tt)$ and $(xx)$ components (first line), and the $(yy)$ and $(zz)$ components (second line):
\begin{eqnarray}
(tt): \; H_1 H_2 + H_1 H_3 + H_2 H_3 = \Lambda + \frac{4 \pi G}{\alpha} \frac{B_z^2}{a_1^2 a_2^2}, \qquad
&& (xx): \; \frac{\ddot a_2}{a_2} + \frac{\ddot a_3}{a_3} + H_2 H_3 = \Lambda - \frac{4 \pi G}{\alpha} \frac{B_z^2}{a_1^2 a_2^2},
\nonumber\\
(yy): \; \frac{\ddot a_1}{a_1} + \frac{\ddot a_3}{a_3} + H_1 H_3 = \Lambda - \frac{4 \pi G}{\alpha} \frac{B_z^2}{a_1^2 a_2^2}, \qquad\quad\quad
&& (zz): \; \frac{\ddot a_1}{a_1} + \frac{\ddot a_2}{a_2} + H_1 H_2 = \Lambda +\frac{4 \pi G}{\alpha} \frac{B_z^2}{a_1^2 a_2^2},
\label{eq_E_comp}
\end{eqnarray}
where a dot denotes differentiation with respect to the cosmological time $t$.

In order to solve the system, it is convenient to reformulate the Einstein equations into those that do not explicitly depend on the source $\rho_M$ given in \eqref{eq_T_M} and those that do. From the structure of the cosmological term and the magnetic field contributions, one can obtain two equations involving only the metric components.
One equation is obtained from the combination $(xx) - (yy)$, which gives
\begin{equation}
\label{eq_a_1}
\frac{\ddot a_1}{a_1} - \frac{\ddot a_2}{a_2} + H_1 H_3 - H_2 H_3 = 0.
\end{equation}
We have the second independent equation from $(tt) - (zz)$:
\begin{equation} 
\label{eq_a_2}
\frac{\ddot a_1}{a_1} + \frac{\ddot a_2}{a_2} + H_1 H_3 + H_2 H_3 = 0.
\end{equation}
The sum of the four Einstein equations yields the third equation that involves only the cosmological term:
\begin{equation}
\label{eq_a_3}
\frac{\ddot a_1}{a_1} + \frac{\ddot a_2}{a_2} + \frac{\ddot a_3}{a_3} + H_1 H_2 + H_1 H_3 + H_2 H_3 = 2 \Lambda.
\end{equation}
These are the equations that do not depend on the source $\rho_M$.
Finally, the combination $(xx) + (yy) + (zz) + 3 (tt)$ leads to the fourth independent equation that depends on the source and will be useful in solving the system:
\begin{equation}
\label{eq_a_4}
\frac{\ddot a_1}{a_1} + \frac{\ddot a_2}{a_2} + \frac{\ddot a_3}{a_3} + 2 \left( H_1 H_2 + H_1 H_3 + H_2 H_3 \right) = 3 \Lambda + \frac{4 \pi G}{\alpha} \frac{B_z^2}{a_1^2 a_2^2}.
\end{equation}
The above equations~\eqref{eq_a_1}--\eqref{eq_a_4} are four independent equations which are equivalent to the original equations in~\eqref{eq_E_comp}.

To solve these equations, it is convenient to parametrize the directional scale factors as~\cite{Casadio:2023ggm}
\begin{eqnarray}
a_1(t) &=& V(t)^{1/3} \, \mathrm{e}^{\gamma(t) + \beta(t)}, 
\nonumber \\
a_2(t) &=& V(t)^{1/3} \, \mathrm{e}^{\gamma(t) - \beta(t)}, 
\nonumber \\
a_3(t) &=& V(t)^{1/3} \, \mathrm{e}^{-2 \gamma(t)}.
\label{dirscal}
\end{eqnarray}
Substituting this parametrization into Eq.~\eqref{eq_a_1} gives a simple first-order relation for the parameter $\beta(t)$ that describes the anisotropy in $x$ and $y$:
\bea
\frac{d}{dt} \left( V \dot\beta \right) = 0,
\eea
which can be integrated to
\bea
\dot \beta(t) = \frac{\beta_0}{V(t)},
\label{eq_V_1}
\eea
where $\beta_0$ is an integration constant.
Interestingly, if $\beta_0 = 0$ for a suitable initial condition, $\beta(t)$ becomes constant, and the universe is axially symmetric, i.e. $a_1 = a_2$ after a suitable rescaling of spatial coordinates~$x$ and $y$. Such a model is known as an axisymmetric model.

It is straightforward to rewrite the rest of equations in terms of $V$ and $\gamma$. From Eq.~\eqref{eq_a_2}, we get
\begin{equation}
\label{eq_V_2}
\frac13 \frac{\ddot V}{V} - \frac13 \frac{\dot V^2}{V^2} + \ddot\gamma + 3 \dot\gamma^2 + \frac{\dot V}{V} \dot \gamma + \dot\beta^2 = 0,
\end{equation}
and Eq.~\eqref{eq_a_3} gives
\begin{equation}
\label{eq_V_3}
\frac{\ddot V}{V} - \frac13 \frac{\dot V^2}{V^2} + 3 \dot\gamma^2 + \dot\beta^2 = 2 \Lambda.
\end{equation}
Finally, the equation with explicit coupling to the magnetic field~\eqref{eq_a_4} becomes
\begin{equation}
\label{eq_V_4}
\frac{\ddot V}{V} = 3 \Lambda + \frac{4 \pi G B_z^2}{\alpha V^{4/3} \mathrm{e}^{4 \gamma}}.
\end{equation}

We therefore have four coupled differential equations~\eqref{eq_V_1} -- \eqref{eq_V_4}, but only three unknown functions $a_1(t), a_2(t)$ and $a_3(t)$. It appears that the system is overdetermined and that there may not be consistent solutions. Actually, the four equations in Eq.~\eqref{eq_E_comp} are not independent. Denoting the differences between the left- and right-hand sides of Eqs.~\eqref{eq_E_comp} by $f_0, f_1, f_2$ and $f_3$, one can verify that, for constant couplings $G, \Lambda$ and $\alpha$, they satisfy
\begin{equation}
\label{eq_ind}
\dot f_0 + \Big(\sum_{i=1}^3 H_i \Big) f_0 = \sum_{i=1}^3 ( H_i f_i ).
\end{equation}
Hence, if three of the independent equations among the original set are satisfied, the remaining one may be satisfied. For example, if we solve $f_0=0$ and two out of $f_i=0$, say $f_1=f_2=0$, then we obtain $H_3 f_3 = 0$ from Eq.~\eqref{eq_ind} since $f_0 = 0$ is a constraint equation and its time derivative automatically vanishes. This gives $f_3 = 0$ unless $H_3 = 0$, which is true in general. On the other hand, if we solve $f_i = 0$ for $i = 1, 2, 3$, then Eq.~\eqref{eq_ind} implies
\begin{equation}
f_0 = \frac{C}{a_1 a_2 a_3},
\end{equation}
where $C$ is an integration constant. Upon imposing the initial condition $f_0=0$, we obtain $f_0=0$. So in general, it is not enough to solve the dynamical equations~$f_1=f_2=f_3=0$ only, but we have to check if $f_0=0$ is satisfied. In this way, there is no consistency problem. A similar observation has been made, for example, in Refs.~\cite{Maeda:2004hu, Akune:2006dg}.

However, if the cosmological and Newton couplings become time-dependent in a quantum theory, we have
\begin{equation}
\dot f_0 + f_0 \sum_{i=1}^3 H_i - \sum_{i=1}^3 (H_i f_i) = - \dot\Lambda(t) - \frac{4 \pi B_z^2 \dot G(t)}{\alpha a_1^2(t) a_2^2(t)}.
\end{equation}
We see that the right-hand side does not vanish, leading to a potential consistency problem. 
When quantum effects come into play, generally we expect that they induce an additional energy density $\rho_q(t)$, modifying the $(tt)$ component of the Einstein equations
\begin{equation}
H_1 H_2 + H_1 H_3 + H_2 H_3 = \Lambda(t) + \frac{4 \pi G(t)}{\alpha} \frac{B_z^2}{a_1^2 a_2^2} + 8 \pi G_0 \rho_q(t).
\end{equation}
Since we are considering radiation-dominated universe with the stress-energy density given by~\eqref{eq_T_M}, it is natural to expect that the quantum corrections to the energy-momentum tensor $T_q{}^\mu{}_\nu$ have the same traceless form. This means that the quantum corrections do not appear in the equations without $\rho_M$ and so the quantum-improved versions of Eqs.~\eqref{eq_V_2} and~\eqref{eq_V_3} remain the same, and that of Eq.~\eqref{eq_V_4} becomes
\begin{equation}
\label{eq_V_4q}
\frac{\ddot V_q(t)}{V_q(t)} = 3 \Lambda(t) + \frac{4 \pi G(t) B_z^2}{\alpha V_q(t)^{4/3} \mathrm{e}^{4 \gamma_q(t)}} + 8 \pi G_0 \rho_q(t).
\end{equation}
Now we have four variables $a_1(t), a_2(t), a_3(t)$ and $\rho_q(t)$, and four equations in the quantum-improved  theory, so the system is consistent.
Technically, one can first solve $V(t)$ and $\gamma(t)$ from Eqs.~\eqref{eq_V_2} and~\eqref{eq_V_3}, and then determine the associated quantum energy density $\rho_q(t)$ from Eq.~\eqref{eq_V_4q}.

Unfortunately, these equations are difficult to solve exactly. So we first derive perturbative late-time solutions for the classical case with constant couplings $G = G_0$ and $\Lambda = \Lambda_0$. We then incorporate quantum improvements by promoting the couplings to running quantities, $G = G(k)$ and $\Lambda = \Lambda(k)$, together with a physically reasonable identification of the cutoff scale $k = k(t)$.

\subsection{Classical Solution at late time for $\Lambda_0 = 0$ case} \label{CBIL0}

\subsubsection{Power series solution}

In this subsection, we solve perturbatively for the volume element $V(t)$ and the anisotropy parameter $\gamma(t)$ from the two coupled second-order differential equations given in~\eqref{eq_V_2} and~\eqref{eq_V_3} for $\Lambda_0 = 0$. Once $V(t)$ is obtained, $\beta(t)$ can be determined from~\eqref{eq_V_1}. Note that in this case the cosmological term vanishes, $\Lambda = 0$, since it is independent of $k$ at the classical level. We also use Eq.~\eqref{eq_V_4} to relate the integration constants to the magnetic field parameter $B_z$.

Let us consider an expanding universe with $\dot V > 0$, in which the volume element diverges as $t \to \infty$. In this limit, the term involving the magnetic field becomes subleading. Equation~\eqref{eq_V_4} then implies that the second derivative of the leading term in $V(t)$ satisfies $\ddot V = 0$ for $\Lambda = 0$. At leading order, this yields the late-time Kasner solution
\begin{equation}
V(t) = V_0 (t - t_0),
\end{equation}
where $V_0$ and $t_0$ are integration constants of the second-order differential equation.

The leading-order solution for the anisotropy parameter $\gamma(t)$ can be obtained from the difference between Eqs.~\eqref{eq_V_2} and~\eqref{eq_V_3}. Setting $\Lambda = 0$ and using the leading-order solution for $V(t)$, we obtain
\bea
\frac{d}{dt} (\dot\gamma V) = 0,
\eea
which can be twice integrated:
\bea
\gamma(t) = \frac{\gamma_0}{V_0} \ln(t - t_0) + \gamma_1,
\eea
where $\gamma_0$ and $\gamma_1$ are integration constants.

Similarly, the solution for $\beta(t)$ contains two integration constants $\beta_0$ in Eq.~\eqref{eq_V_1} and another one after we integrate it. In total, there are six integration constants, as expected for three second-order differential equations governing the scale factors. However, the system also contains an additional equation that acts as a constraint rather than an independent evolution equation. This constraint restricts the allowed initial conditions and imposes a relation among the integration constants, fixing one of them in terms of the others and leaving five independent constants.

Substituting the solutions for $V(t)$ and $\gamma(t)$ into Eq.~\eqref{eq_V_3} yields a relation that constrains $V_0$ in terms of the other constants $\gamma_0$ and $\beta_0$:
\begin{equation}
\label{eq_constr1}
V_0 = \sqrt{9 \gamma_0^2 + 3 \beta_0^2}.
\end{equation}
Consequently, only two of these constants remain independent.

A more physically motivated approach to obtaining the next-order term, which arises from the magnetic field contribution, is to substitute the leading-order solution into the magnetic-field term. In this way, Eq.~\eqref{eq_V_4} yields the following correction to the volume element:
\begin{equation}
V(t)= V_0 (t - t_0) + \frac{C}{(2/3 - s) (5/3 - s)} (t - t_0)^{5/3 - s} + \cdots,
\end{equation}
where
\begin{equation}
C = \frac{4 \pi G_0 B_z^2}{\alpha V_0^{1/3} \mathrm{e}^{4 \gamma_1}},
\qquad
s = \frac{4 \gamma_0}{V_0}.
\end{equation}
The subleading correction to $\gamma(t)$ can be obtained as
\begin{equation}
\gamma(t) = \gamma_1 + \frac{s}{4} \ln(t - t_0) + \frac{C}{(2/3 - s) (5/3 - s)} \left( \frac{40 - 33 s}{12 (2 - 3 s) V_0} \right) (t - t_0)^{2/3 - s} + \cdots.
\end{equation}

It is technically more convenient to determine the power-series solutions for $V(t)$ and $\gamma(t)$ directly from the coupled equations~\eqref{eq_V_2} and~\eqref{eq_V_3}. The resulting expansions are
\begin{eqnarray}
V(t) &=& V_0 (t - t_0) + V_1 (t - t_0)^{5/3 - s} - \frac{(5 - 3 s) V_1^2}{(2 - 3 s) V_0} (t - t_0)^{7/3 - 2 s} + \frac{3 (5 - 3 s) (4 - 3 s) V_1^3}{2 (2 - 3 s)^2 V_0^2} (t - t_0)^{3 - 3 s}
\nonumber\\
&-& \frac{(5 - 3 s) (7 - 6 s) (17 - 12 s) V_1^4}{3 (2 - 3 s)^3 V_0^3} (t - t_0)^{11/3 - 4 s} + \cdots.
\label{eq_power_V}
\end{eqnarray}
and 
\begin{eqnarray}
\gamma(t) &=& \gamma_1 + \frac{s}{4} \ln(t - t_0) + \frac{(40 - 33 s) V_1}{12 (2 - 3 s) V_0} (t - t_0)^{2/3 - s} - \frac{3 (40 - 71 s + 30 s^2) V_1^2}{8 (2 - 3 s)^2 V_0^2} (t - t_0)^{4/3 - 2 s} 
\nonumber\\
&+& \frac{(6560 - 17748 s + 15525 s^2 - 4401 s^3) V_1^3}{72 (2 - 3 s)^3 V_0^3} (t - t_0)^{2 - 3 s} + \cdots.
\label{eq_power_gamma}
\end{eqnarray}
Equation~\eqref{eq_V_4} then determines the coefficient $V_1$ as
\begin{equation}
V_1 = \frac{4 \pi G_0 B_z^2}{(2/3 - s) (5/3 - s) \alpha V_0^{1/3} \mathrm{e}^{4 \gamma_1}}. 
\label{eq_def_V1}
\end{equation}
The other anisotropy parameter $\beta(t)$ is determined from Eq.~\eqref{eq_V_1} as
\begin{equation}
\beta(t) = \int_{t_0}^t \frac{\beta_0}{V(t)}dt.
\label{anistropyxy}
\end{equation}
Given the solution for $V(t)$ in Eq.~\eqref{eq_power_V}, this integral can be evaluated straightforwardly, and we do not present the explicit expression here.

The parameters $V_0, \gamma_0$ and $\beta_0$ must satisfy the constraint given by Eq.~\eqref{eq_constr1} which is transformed into
\begin{equation}
\label{eq_relation1}
s^2 = \frac{16}{9} \frac{V_0^2 - 3 \beta_0^2}{V_0^2}.
\end{equation}
This relation imposes an upper bound on $s$, while the requirement that the powers of $t$ in the higher-order terms decrease provides the lower bound
\begin{equation} \label{eq_s_range}
\frac23 < s \le \frac43.
\end{equation}
The upper (lower) bound corresponds to $\beta_0/V_0 = 0$ ($\beta_0/V_0 = 1/2$). Note that within this parameter region, $V_1 < 0$.

At this point, we can see that the integration constant $\gamma_1$ merely reflects a rescaling of the spatial coordinates as $(x', y', z') = (\mathrm{e}^{\gamma_1} x,  \mathrm{e}^{\gamma_1} y, \mathrm{e}^{-2\gamma_1} z)$, under which the magnetic field transforms as $B_z' = \mathrm{e}^{-2 \gamma_1} B_z$.
To make the dimension clearer, let us write $\gamma_1 = (s/4) \ln V_0'$, so that, when combined with the next $\ln(t - t_0)$ term, the argument of the logarithm in the resulting expression $(s/4)\ln(V_0' (t - t_0))$ is dimensionless ($V_0'$ has the dimension of inverse time).

Note that for $\beta_0 = 0$, Eq.~\eqref{eq_relation1} gives $s = 4/3$, while Eq.~\eqref{eq_V_1} implies $\beta(t)$ is a constant, which can be absorbed into a redefinition of the coordinates $x$ and $y$, analogous to the parameter $\gamma_1$. This corresponds to an axially symmetric solution with $a_1 = a_2$:
\begin{eqnarray}
V(t) &=& V_0 ( t - t_0) + V_1 (t - t_0)^{1/3} + \frac{V_1^2}{2 V_0} (t - t_0)^{-1/3} - \frac{V_1^4}{24 V_0^3} (t - t_0)^{-5/3} + \cdots , \qquad
V_1 = - \frac{18 \pi G B_z^2}{\alpha V_0^{1/3}},
\label{eq_symm_V}
\\
\gamma(t) &=& \frac13 \ln(V_0' (t - t_0)) + \frac{V_1}{6 V_0} (t - t_0)^{-2/3} + \frac{V_1^2}{8 V_0^2} (t - t_0)^{-4/3} - \frac{V_1^3}{9 V_0^3} (t - t_0)^{-2} + \frac{17 V_1^4}{576 V_0^4} (t - t_0)^{-8/3} + \cdots.\qquad
\label{eq_symm_gamma}
\end{eqnarray}

\subsubsection{anisotropy}

Now we can check if the isotropy is achieved as time goes on. First, let us examine this for the anisotropy parameter $\beta(t)$. We see from the solution~\eqref{anistropyxy} that if $\beta_0=0$, $\beta(t)$ is a constant. We can then redefine the coordinates $x$ and $y$ to absorb the difference between these coordinates, and so isotropy in these directions is realized. 
This is the solution given in \eqref{eq_symm_V} and \eqref{eq_symm_gamma} and is called the axisymmetric model.
On the other hand, if $\beta_0 \ne 0$, it is clear that the realization of isotropy depends on whether Eq.~\eqref{anistropyxy} is finite or not in the infinite time limit. This is determined by the leading behaviour of $V(t)$ given in Eq.~\eqref{eq_power_V}. Since its leading behaviour is linear in $t$, the integral is divergent. We thus conclude that the isotropy is not realized for general initial conditions ($\beta_0 \ne 0$). This gives a Kasner-type universe~\cite{Kasher:1921zz}.
Still the fact that there is a possibility that the isotropy may be realized for $\beta_0=0$ is consistent with the applied magnetic field since it is only in the $z$ direction but is homogeneous in the $x$ and $y$ directions, and whether it is realized or not is determined by the initial condition.

How about another anisotropy parameter $\gamma(t)$?
We see from the solution~\eqref{eq_power_gamma} that its leading behavior is logarithmic ($\gamma_1$ is absorbed into the logarithmic term) and its coefficient would not vanish unless $s=0$. However, it does not vanish 
according to Eq.~\eqref{eq_s_range}. So isotropy in $z$ direction is not realized.

\subsection{Quantum-improved Solution for $\Lambda_0 = 0$ case} \label{QBIL0}

For the case $\Lambda_0 = 0$, we use the average Hubble parameter of quantum-improved solution as a physically reasonable identification of the cutoff scale~\cite{Chen:2024ebb, Chen:2025ybu}:
\begin{equation}
k = \xi \frac{\dot V_q}{V_q}.
\end{equation}
With this identification together with the $k$-dependence of $G$ and $\Lambda$ in Eqs.~\eqref{gravterms}, the running cosmological term and Newton coupling become time-dependent:
\begin{eqnarray}
\Lambda(t) &=& G_0 \tilde\nu \frac{\dot V_q^4}{V_q^4} + G_0^2 \tilde\nu_1 \frac{\dot V_q^6}{V_q^6} + \cdots,
\\
G(t) &=& G_0 \left( 1 - G_0 \tilde\omega \frac{\dot V_q^2}{V_q^2} + G_0^2 \tilde\omega_1 \frac{\dot V_q^4}{V_q^4} + \cdots \right),
\end{eqnarray}
where $\tilde\nu = \xi^4 \nu, \tilde\nu_1 = \xi^6 \nu_1, \tilde\omega = \xi^2 \omega, \tilde\omega_1 = \xi^4 \omega_1, \cdots$. 
Following the same procedure as in the classical case, we obtain the quantum-corrected power-series solutions $V_q(t)$ and $\gamma_q(t)$ from the coupled equations~\eqref{eq_V_2} and~\eqref{eq_V_3}, now including the running cosmological term. The corresponding leading-order quantum corrections, $\delta V = V_q - V, \delta\gamma = \gamma_q - \gamma$, are
\begin{eqnarray}
\label{quantumv1}
\delta V(t) &\!\!=\!\!& \frac{3 (3 s - 4) V_0 G_0 \tilde\nu}{2 (3 s - 8)} (t - t_0)^{-1} - \frac{(81 s^4 - 459 s^3 + 1728 s^2 - 2112 s + 544) G_0 \tilde\nu V_1}{ 4(3 s - 8) (3 s + 1) (3 s + 4)} (t - t_0)^{-1/3 - s} + \cdots,
\\
\delta \gamma(t) &\!\!=\!\!& \frac{(9 s^2 \!-\! 12 s \!+\! 32) G_0 \tilde\nu}{16 (3 s - 8)} (t \!-\! t_0)^{-2} \!-\! \frac{(891 s^4 \!-\! 3753 s^3 \!+\! 11016 s^2 \!-\! 2640 s \!-\! 3712) G_0 \tilde\nu V_1}{48 (3 s - 8) (3 s + 1) (3 s + 4) V_0} (t \!-\! t_0)^{-4/3 - s} \!+\! \cdots. \qquad
\label{quantumv2}
\end{eqnarray}

\noindent
{\bf Physical meaning of the quantum effects:}

The leading term in the quantum corrections exists even without the magnetic field and is parametrized by $V_0$, while the next-order corrections are present due to the magnetic field, parametrized by $V_1$.
For the range~\eqref{eq_s_range} of $s$, the first correction by the quantum effects is positive, so the universe expands more rapidly than without quantum effects. As $V_q(t) > V(t)$, the parameter $\beta(t)$, which represents the anisotropy between $a_1$ and $a_2$ (see Eq.~\eqref{dirscal} and \eqref{anistropyxy}), in the quantum case is smaller than in the classical case. The parameter $\gamma(t)$, which represents the anisotropy between the longitudinal and transverse directions, receives a negative quantum contribution. Thus the ratios $a_1/a_3$ and $a_2/a_3$ decrease in the presence of quantum corrections compared to the classical case, reducing the relative difference between directional scale factors. Although these quantum effects are subdominant, they reduce the anisotropy during the intermediate stage; the solution continues to the Kasner-type behavior at late times.

The quantum correction to the energy density is given as
\begin{equation}
\rho_{q}(t) = \frac{3 \tilde\nu}{2 \pi (3 s - 8)} t^{-4} + \frac{(81 s^4 - 189 s^3 - 162 s^2 + 1668 s - 1216) V_1 \tilde\nu}{144 \pi (3 s - 8) V_0} t^{-10/3-s} + \cdots.
\end{equation}
Note that the first quantum term is negative.


\subsection{Estimation of the effective magnetic field strength}

In this subsection, we examine the approximate initial magnetic field strength in the presence of quantum corrections within an anisotropic universe. Although the assumption of a spatially homogeneous magnetic field aligned along a single direction in a BI spacetime is highly idealized, it nevertheless provides a useful framework to probe potential effects arising from quantum corrections to the gravitational coupling and cosmological term, as suggested in~\cite{Casadio:2023ggm}.

To clarify this point, we define the effective magnetic field strength $\mathfrak{B}$ by
\begin{equation}
\mathfrak{B}^2 = \frac{1}{2} F^{\mu \nu} F_{\mu \nu} = \frac{B_z^2}{a_1^2 a_2^2}~. 
\end{equation}
which is a coordinate-invariant quantity.
Observational constraints, as reported in~\cite{Barrow:1997, Vachaspati:2021}, typically bound the current cosmological magnetic fields to be no larger than approximately $10^{-9}\,\text{G}$ under the assumption of coherence over megaparsec (Mpc) scales or larger. Taking this as the present-day value, we investigate how quantum corrections modify the corresponding magnetic field strength relative to the classical anisotropic scenario.

In the isotropic limit, as discussed in~\cite{Casadio:2023ggm}, the inferred initial magnetic field becomes extremely large. This is due to the large number of e-folds, $N_e = 60$, corresponding to a redshift $z \sim 10^{50}$. Since the magnetic field scales as $\mathfrak{B} \propto (a_1 a_2)^{-1}$, this leads to a significant enhancement of its initial value.
In the classical anisotropic case, the product $a_1 a_2$ in late time can be estimated using~\eqref{dirscal} together with the leading-order classical solutions for $V(t)$ and $\gamma(t)$, yielding
\begin{equation}
a_1 a_2 \sim V^{2/3} \mathrm{e}^{2 \gamma(t)} \sim t^{2/3 + s/2}.
\end{equation}
The allowed range of $s$, given in Eq.~\eqref{eq_s_range} as $2/3 < s \le 4/3$, implies that at late times the magnetic field decays as $\mathfrak{B} \propto (a_1 a_2)^{-1} \propto t^{- (2/3 + s/2)}$ with the decay exponent lying between $1$ and $4/3$. This indicates that, within this Kasner era, evolving backwards in time leads to a significantly enhanced initial magnetic field, similar to the isotropic case. We also observe that, in the limit of vanishing anisotropy parameter $\beta$, the magnetic field decays more rapidly than in a non-axisymmetric universe since $s$ is largest for $\beta_0=0$ (see Eq.~\eqref{eq_relation1}).
Another important point is that, although quantum corrections are small at late times, the leading quantum contribution in~\eqref{quantumv1} is positive and that from $\gamma$ is subdominant. So the quantum-improved volume~$V_{q}(t)$ would be larger than the classical volume~$V(t)$ at a given time. This leads to a faster growth of the product $a_1 a_2$, and consequently to a quicker decay of the magnetic field in the presence of quantum corrections.

As discussed above, during cosmic expansion, magnetic fields are diluted, implying that the presently observed small field strength generally requires a large initial value, at least in an isotropic universe. In Ref.~\cite{Casadio:2023ggm}, it was argued that anisotropic expansion can alleviate this requirement, since at early times the Kasner regime differs from the late-time one. In particular, a faster expansion along the magnetic-field direction, combined with slower expansion (or contraction) in the transverse directions, can significantly suppress the decay of the field.

In our case, however, our solutions~\eqref{eq_power_V} and \eqref{eq_power_gamma} indicate that the expansion in the transverse directions $a_1(t)a_2(t)$ is dominant whereas the expansion in $a_3(t)$ is suppressed by $\gamma(t)$. Furthermore, the leading quantum contributions enhance this tendency, leading to the dilution of the magnetic field at late times. Therefore, any initial suppression of the decay is likely to be temporary and does not qualitatively alter the overall trend. As a result, the initial magnetic field is expected to be sufficiently larger to remain compatible with current observational bounds.

\subsection{Classical Solution at late time for $\Lambda = \Lambda_0 > 0$ case}

\subsubsection{Power series solution}

In this subsection, we discuss the classical solution for $\Lambda_0 > 0$. In the absence of the magnetic field, Eqs.~\eqref{eq_V_4} and~\eqref{eq_V_3} tell us that the leading solution would be
\bea
V \sim \mathrm{e}^{\sqrt{3 \Lambda_0} \, t}, \qquad
\dot \gamma \sim \mathrm{e}^{-\sqrt{3\Lambda_0}\, t}.
\label{classical2}
\eea
There is another possibility $V \sim \exp(-\sqrt{3\Lambda_0}\, t)$, but we do not consider this solution since we are interested in the expanding universe.
So it is natural to introduce a new ``time'' coordinate 
\begin{equation}
\tau \equiv \frac1{\sqrt{3 \Lambda_0}} \mathrm{e}^{\sqrt{3 \Lambda_0} \, t}.
\end{equation}
Note that translational symmetry $t \to t - t_0$ reduces to scaling symmetry $\tau \to \exp\left( -\sqrt{3\Lambda_0} \, t_0 \right) \tau$.
In terms of $\tau$, Eqs.~\eqref{eq_V_2} and~\eqref{eq_V_3} become
\begin{eqnarray} 
&& \tau^2 \frac{V''}{V} + \tau \frac{V'}{V} - \frac{\tau^2}3 \frac{V'^2}{V^2} + 3 \tau^2 \gamma'^2 + \frac{\tilde\beta_0^2}{V^2} = \frac{2 \Lambda}{3 \Lambda_0},
\\
&& \frac{\tau^2}3 \frac{V''}{V} + \frac{\tau}3 \frac{V'}{V} - \frac{\tau^2}3 \frac{V'^2}{V^2} + \tau^2 \gamma'' + \tau \gamma' + 3 \tau^2 \gamma'^2 + \tau^2 \gamma' \frac{V'}{V} + \frac{\tilde\beta_0^2}{V^2} = 0,
\end{eqnarray}
where primes denote derivatives with respect to $\tau$, and $\tilde\beta_0 = \beta_0/\sqrt{3 \Lambda_0}$. The classical solutions ($\Lambda = \Lambda_0$) admit the asymptotic expansions
\begin{eqnarray}
V(\tau) &=& U_0 \tau \left( 1 + U_1 \tau^{-4/3} - \left( \frac{3 \tilde\beta_0^2}{4 U_0^2}  + \frac{9 \delta_2^2}{4} \right) \tau^{-2}  + \frac{32 U_1 \delta_2}{7} \tau^{-7/3}  - \frac{5 U_1^2}{2} \tau^{-8/3} - \left( \frac{\tilde\beta_0^2 U_1}{20 U_0^2} + \frac{7 U_1 \delta_2^2}{4}  \right) \tau^{-10/3} \right.
\nonumber\\
&&  + \frac{256 U_1^2 \delta_2}{77} \tau^{-11/3}  - \frac{3 U_1^3}{2}\tau^{-4}  + \left( \frac{400 U_1 \delta_2^3}{273} + \frac{16 \tilde\beta_0^2 U_1 \delta_2}{91 U_0^2} \right) \tau^{-13/3}  - \left( \frac{377 \tilde\beta_0^2 U_1^2}{1960 U_0^2} + \frac{1903 U_1^2 \delta_2^2}{392}  \right) \tau^{-14/3}
\nonumber\\
&& \left.  + \frac{384 U_1^3 \delta_2}{77} \tau^{-5}  - \left( \frac{13 U_1^4}{8} + \frac{\tilde\beta_0^4 U_1}{160 U_0^4} + \frac{143 U_1 \delta_2^4}{96} + \frac{27 U_1 \tilde\beta_0^2 \delta_2^2}{80 U_0^2}  \right) \tau^{-16/3} + \cdots \right),
\label{vexp}
\\
\gamma(\tau) &=& \delta_1 +  \delta_2 \tau^{-1} - \frac{4 U_1}{3} \tau^{-4/3} + \frac{U_1 \delta_2}{3} \tau^{-7/3}  +  \left( \frac{3 \delta_2^3}{4} + \frac{\tilde\beta_0^2 \delta_2}{4 U_0^2} \right) \tau^{-3}  - \left( \frac{44 \tilde\beta_0^2 U_1}{105 U_0^2}  + \frac{68 U_1 \delta_2^2}{21}   \right) \tau^{-10/3}
\nonumber\\
&&  + \frac{57 U_1^2 \delta_2}{14} \tau^{-11/3} - \frac{16 U_1^3}{9} \tau^{-4} +  \left( \frac{35 U_1 \delta_2^3}{36} + \frac{43 \tilde\beta_0^2 U_1 \delta_2}{780 U_0^2} \right) \tau^{-13/3}  + \left( \frac{72 \tilde\beta_0^2 U_1^2}{385 U_0^2} - \frac{24 U_1^2 \delta_2^2}{11}  \right) \tau^{-14/3} + \cdots,
\label{gammaexp}
\end{eqnarray}
where $U_0$, $\delta_1$ and $\delta_2$ are integration constants and $U_1$ is defined as
\bea
U_1 = -\frac{3 \pi G_0 B_z^2}{2 \alpha \Lambda_0 U_0^{4/3} \mathrm{e}^{4\delta_1}}.
\eea
$\beta(t)$ is again given by~\eqref{anistropyxy}, and we refrain from giving it explicitly.

The fractional power comes as the effect of the magnetic field, since the terms carry fractional powers of $V$ (see Eq.~\eqref{eq_V_4}). Note also that the term $\delta_2 \tau^{-1}$ in $\gamma(\tau)$ reproduces the exponentially decaying behavior expected from Eq.~\eqref{classical2}. Here, we observe that the classical magnetic field strength decays exponentially as $\mathfrak{B} \propto \exp(-\frac{2}{3}\sqrt{3\Lambda_0}, t)$, and that the cosmological constant acts as the primary driving force governing the late-time evolution toward an isotropically expanding de Sitter universe, as expected.

\subsubsection{anisotropy}

Let us check if the isotropy is achieved as the time goes. First, we examine this for the anisotropy parameter $\beta(t)$. We can see from the solution~\eqref{anistropyxy} again that if $\beta_0=0$, $\beta(t)$ is a constant and so isotropy in $x$ and $y$ directions is realized. On the other hand, if $\beta_0 \ne 0$, Eq.~\eqref{anistropyxy} is finite in the infinite time limit since the leading behavior of $V(t)$ is exponential as given in Eq.~\eqref{vexp}.
So the isotropy in the $xy$ directions is realized eventually after the possible redefinitions of $x$ and $y$ for general initial conditions.

As to the other anisotropy parameter $\gamma(t)$,
we see from the solution~\eqref{gammaexp} that it tends to a constant. This could be absorbed into the redefinition of the coordinates.

So we conclude in this case that isotropy is realized in the future.

\subsection{Quantum-improved Solution for $\Lambda_0 > 0$ case}

We again use the identification of the cutoff scale in the presence of a nonvanishing cosmological term $\Lambda_0$, based on the generalization of $k \sim 1/\tau$, proposed in Refs.~\cite{Chen:2024ebb, Chen:2025ybu}:
\begin{equation}
k = \xi \frac{\sqrt{3 \Lambda_0}}{V_q}.
\end{equation}
With this identification, the running cosmological term becomes
\begin{equation}
\Lambda(\tau) = \Lambda_0 \left( 1 - \frac{G_0 \tilde\mu}{V_q^2} + \frac{G_0^2 \tilde\mu_1 + G_0 \tilde\nu/\Lambda_0}{V_q^4} + \cdots \right), 
\end{equation}
where the rescaled parameters are defined by
$\tilde\mu = 3 \Lambda_0 \xi^2 \mu, \, \tilde\mu_1 = (3 \Lambda_0)^2 \xi^4 \mu_1, \, \tilde\nu = (3 \Lambda_0)^2 \xi^4 \nu$.
From this expression, one can straightforwardly derive the corresponding leading order quantum corrections:
\begin{eqnarray}
\delta V(\tau) &=& U_0 \tau \left( - \frac{G_0 \tilde\mu}{2 U_0^2} \tau^{-2}  - \frac{4 \delta_2 G_0 \tilde\mu}{5 U_0^2} \tau^{-3}  + \frac{31 U_1 G_0 \tilde\mu}{30 U_0^2} \tau^{-10/3} - \left( \frac{G_0^2 \tilde\mu^2}{8 U_0^4} - \frac{G_0^2 \tilde\mu_1}{16 U_0^4} - \frac{G_0 \tilde\nu}{16 U_0^4 \Lambda_0} - \frac{9 \delta_2^2 G_0 \tilde\mu}{20 U_0^2}  \right) \tau^{-4} + \cdots \right),
\\
\delta \gamma(\tau) &=& \frac{G_0 \tilde\mu}{3 U_0^2} \tau^{-2}  - \frac{\delta_2 G_0 \tilde\mu}{10 U_0^2} \tau^{-3}  - \frac{10 U_1 G_0 \tilde\mu}{63 U_0^2} \tau^{-10/3} + \left( \frac{G_0^2 \tilde\mu^2}{18 U_0^4} + \frac{\tilde\beta_0^2 G_0 \tilde\mu}{6 U_0^4} - \frac{G_0^2 \tilde\mu_1}{36 U_0^4} - \frac{G_0 \tilde\nu}{36 U_0^4 \Lambda_0}  + \frac{9 \delta_2^2 G_0 \tilde\mu}{10 U_0^2}  \right) \tau^{-4} + \cdots
\end{eqnarray}

It is interesting to note that the leading quantum correction to the total volume $V_q(\tau)$ is negative, in contrast to the power expanding solution. This leads to slower expansion of the universe. We find that both anisotropy parameters $\beta(t)$ and $\gamma(t)$ are larger in the presence of quantum corrections than in the classical case. This indicates that quantum corrections slightly enhance the anisotropy during the intermediate stages. However, since the quantum corrections decay exponentially, they do not affect the late-time dynamics. Moreover, the quantum magnetic field strength is also slightly larger than in the classical case, as $V_q < V$.

To obtain the corresponding quantum energy density, we first rewrite~\eqref{eq_V_4q} in terms of $\tau$: 
\begin{equation}
3 \Lambda_0 \tau^2 \frac{V_q''}{V_q} + 3 \Lambda_0 \tau \frac{V_q'}{V_q} = 3 \Lambda + \frac{4 \pi G B_z^2}{\alpha V_q^{4/3} \mathrm{e}^{4 \gamma_q}} + 8 \pi G_0 \rho_q.
\end{equation}
The running Newton coupling is given by
\begin{equation}
G(\tau) = G_0 \left( 1 - \frac{G_0 \tilde\omega}{V_q^2} + \frac{G_0^2 \tilde\omega_1}{V_q^4} + \cdots \right), \end{equation}
where $\tilde\omega = 3 \Lambda_0 \xi^2 \omega$ and $\tilde\omega_1 = (3 \Lambda_0)^2 \xi^4 \omega_1$. 
A straightforward calculation then yields the leading contributions to the quantum energy density
\begin{equation}
\rho_q(\tau) = \frac{3 \Lambda_0 \tilde\mu}{8 \pi U_0^2} \tau^{-2}  - \frac{9 \Lambda_0 \delta_2 \tilde{\mu}}{10 \pi U_{0}^2} \tau^{-3} + \frac{\Lambda_0 U_1 (7 \tilde\mu - 4 \tilde\omega)}{12 \pi U_0^2} \tau^{-10/3} + \cdots. 
\end{equation}
The first term arises from the vacuum solution, the second term is associated with the $\delta_2$ mode, while the third term originates from the magnetic field.



\section{BI universe with Electric Field}
\label{BIele}

We next consider an anisotropic cosmological background of BI type~\eqref{Bianchi_I}, coupled to a homogeneous electric field aligned along the $z$-direction:
\begin{equation}
 F_{03}  = E_z,
\end{equation}
which can be time-dependent or constant.
For this configuration, the energy-momentum tensor is diagonal and given by
\begin{equation}
\label{eq_x}
T_E{}^{\mu}{}_{\nu} = \mathrm{diag}(-\rho_E, \rho_E, \rho_E, -\rho_E),
\qquad
\rho_E = \frac{E_z^2}{2 \alpha a_3^2}.
\end{equation}

The Einstein equations reduce to the following set of equations, corresponding respectively to the $(tt)$ and $(xx)$ components (first line), and the $(yy)$ and $(zz)$ components (second line):
\begin{eqnarray}
\label{eq_Einstein_E}
(tt): \; H_1 H_2 + H_1 H_3 + H_2 H_3 = \Lambda + \frac{4 \pi G}{\alpha} \frac{E_z^2}{a_3^2},
\qquad
&& (xx): \; \frac{\ddot a_2}{a_2} + \frac{\ddot a_3}{a_3} + H_2 H_3 = \Lambda - \frac{4 \pi G}{\alpha} \frac{E_z^2}{a_3^2},
\nonumber\\
(yy): \; \frac{\ddot a_1}{a_1} + \frac{\ddot a_3}{a_3} + H_1 H_3 = \Lambda - \frac{4 \pi G}{\alpha} \frac{E_z^2}{a_3^2},
\qquad
&& (zz): \; \frac{\ddot a_1}{a_1} + \frac{\ddot a_2}{a_2} + H_1 H_2 = \Lambda + \frac{4 \pi G}{\alpha} \frac{E_z^2}{a_3^2},
\end{eqnarray}
and the Maxwell equation becomes
\begin{equation}
\label{eq_Maxwell_E}
\dot E_z + E_z (H_1 + H_2 - H_3) = 0.
\end{equation}

Now let us make an important observation.
It is known that the electric and magnetic fields are related by a Hodge dual transformation. In a BI cosmological background, we consider an electric field oriented along the $z$-direction:
\bea
F_{[2]} = E_z(t) dt \wedge dz = \frac{E_z}{a_3} \vartheta^{\bar 0} \wedge \vartheta^{\bar 3},
\eea
where $\vartheta^{\bar \mu}$ denotes the orthonormal frame. Taking the Hodge dual,
\bea
F^*_{\bar\alpha \bar\beta} = \frac1{2!} \epsilon_{\bar\alpha \bar\beta \bar\mu \bar\nu} F^{\bar\mu \bar\nu}, \qquad \epsilon_{\bar0 \bar1 \bar2 \bar3} = 1,
\eea
one finds
\bea
F^*_{\bar1 \bar2} = - \frac{E_z}{a_3},
\eea
which means that
\bea
- B_z = F^*_{12} = - \frac{E_z a_1 a_2}{a_3}.
\eea
The Maxwell equation~\eqref{eq_Maxwell_E} then tells us that
\bea
\frac{d}{dt} \ln E_z + \frac{d}{dt} \left( \ln a_1 + \ln a_2 - \ln a_3 \right) = 0,
\eea
giving
\begin{equation}
E_z \frac{a_1 a_2}{a_3} = \mathrm{const.}
\end{equation}
If we express the integration constant as $B_z$, Eqs.~\eqref{eq_Einstein_E} reduce exactly to those obtained for the magnetic case in~\eqref{eq_E_comp}. This demonstrates the equivalence between the electric and magnetic descriptions under Hodge duality. Consequently, it is not necessary to solve the equations; the corresponding solutions can be obtained directly via this duality.


\section{Summary}
\label{summary}

In this paper, we first studied the late-time cosmological evolution of an anisotropic BI universe in the presence of a homogeneous isotropic perfect fluid, incorporating quantum corrections to the gravitational coupling and the cosmological term within the asymptotic safety framework. We have then considered the configuration in which a magnetic field is aligned along a single spatial direction as a typical concrete example of the universe filled with radiation. The main difference between these cases is whether the background radiation is isotropic or anisotropic, but both cases have traceless energy-momentum tensors.
We have examined the classical solutions thoroughly and studied how the BI universe is influenced by quantum corrections at late times for two distinct branches of renormalization group trajectories, corresponding to $\Lambda_0 = 0$ and $\Lambda_0 > 0$.

For the radiation-dominated case with a perfect fluid of homogeneous radiation, we have found that the classical evolution of the BI universe with vanishing cosmological term with the total volume exhibits a nontrivial dependence on the anisotropy parameter through logarithmic corrections, which are absent for other types of matter. These corrections reflect the persistent influence of anisotropy and lead to a slower approach toward isotropy compared to other cosmological epochs.
When quantum effects are incorporated through scale-dependent gravitational and cosmological couplings, we observe that the leading quantum correction matches the structure of higher-order terms, in particular the $\kappa^4$ term in the classical anisotropy expansion. This correspondence indicates that the running cosmological term $\Lambda(k)$ effectively generates contributions of comparable magnitude to higher-order anisotropy effects. In particular, the $(t - t_0)^{-1/2}$ dependence of the leading quantum term modifies the late-time behavior of the volume element during the intermediate stage, and the quantum corrections lead to faster isotropization compared to the purely classical case. 

We have also studied the evolution of the BI universe when we have the anisotropic magnetic fields in $z$ direction with again traceless energy-momentum tensor. We have observed that the Einstein–Maxwell system becomes overdetermined when these couplings acquire time dependence, leading to a consistency problem. This is in contrast to the classical case, where, out of the four equations, three are independent, and the remaining one serves as a constraint. To resolve this quantum case, we have introduced an additional quantum-induced energy density, modeled as a traceless effective quantum fluid similar to the electromagnetic energy-momentum tensor. With this,
we have derived the quantum-improved power-series solutions for the volume element and the anisotropy parameter in terms of cosmic time for both $\Lambda_0=0$ and $\Lambda_0>0$.

For $\Lambda_0=0$, the classical universe evolves towards a Kasner-type regime, where the subleading behavior is governed by the anisotropy parameter $s$. We have obtained constraints on $s$ to ensure that higher-order terms remain subdominant at late times. In this case, anisotropy persists and isotropization is not generically achieved, except under special initial conditions in the axisymmetric configuration. The magnetic field decays as a power law determined by the anisotropic expansion, implying that its initial value must be significantly larger to match present observational bounds. The inclusion of quantum corrections modifies the late-time behavior of both the volume and anisotropy parameters, enhancing the expansion rate compared to the classical case and thereby accelerating the dilution of the magnetic field.

For $\Lambda_0>0$, the late-time dynamics are dominated by the cosmological term, and the universe asymptotically approaches an isotropic de Sitter phase, consistent with the cosmic no-hair theorem. In this regime, the magnetic field decays exponentially, and anisotropies are effectively washed out. Overall, our analysis highlights how the interplay between electromagnetic fields and quantum gravitational effects influences the evolution of anisotropic cosmologies, particularly at late times.

Another notable result is that the additional quantum-induced energy density required for consistency of the field equations is negative for the $\Lambda_0=0$ case, but positive for $\Lambda_0>0$. This indicates that the quantum contribution can act in opposition to the classical behavior, a feature that we also observed in our earlier work for the FLRW universe with a perfect fluid~\cite{Chen:2024ebb}. In the present case, for $\Lambda_{0}=0$, the quantum-corrected volume element is larger than the classical one, even though the universe is in a decelerating phase. In contrast, for $\Lambda_0>0$, the quantum-corrected volume decreases at late times, contrary to the classical expectation of accelerated expansion.

Finally we have noted the Hodge duality which makes the electric backgrounds equivalent to the magnetic ones. So our discussions on the universe with magnetic fields may be directly transformed to the one filled with electric fields.

The present work has focused on the infrared regime of the renormalization group flow, where the running gravitational couplings can be reliably approximated by their late-time expansions. An interesting future direction would be to investigate anisotropic magnetized cosmologies in the ultraviolet fixed-point regime of asymptotically safe gravity, where the scaling behavior of the couplings differs significantly from the infrared limit~\cite{Bonanno:2001xi, Bonnano:2002plb, BF_2017,Gubitosi:2018gsl}. Since primordial magnetic fields and magnetogenesis scenarios are closely connected to the dynamics of the early Universe, it would be interesting to explore whether ultraviolet asymptotic safety effects can influence the generation, amplification, or evolution of cosmological magnetic fields.

%
%
%
%
%

\acknowledgments

C.M.C. and N.O. would like to thank Institute of Fundamental Physics and Quantum Technology, Ningbo University for the hospitality where part of this work was done.
The work of C.M.C. was supported by the National Science Council of the R.O.C. (Taiwan) under the grants NSTC 114-2112-M-008-010 and 115-2918-I-008-006. 
The work of R.M. was supported by the National Science and Technology Council of the R.O.C. (Taiwan) under the grant NSTC 114-2811-M-008-024.

\begin{appendix}

\end{appendix}

\end{document}